# Super-Resolution Ultrasound Localization Microscopy Based on a High Frame-rate Clinical Ultrasound Scanner: An In-human Feasibility Study

Chengwu Huang[1,§], Wei Zhang[2,§], Ping Gong[1], U-Wai Lok[1], Shanshan Tang[1], Tinghui Yin[2], Xirui Zhang[3], Lei Zhu[3], Maodong Sang[3], Pengfei Song[4, 5], Rongqin Zheng[2,*], Shigao Chen[1,*]

[1] Department of Radiology, Mayo Clinic College of Medicine and Science, Mayo Clinic, Rochester, MN, USA
[2] Department of Ultrasound, Guangdong Key Laboratory of Liver Disease Research, Third Affiliated Hospital of Sun Yat-Sen University, Guangzhou, Guangdong, China
[3] Shenzhen Mindray Bio-Medical Electronics Co. Ltd. Shenzhen, Guangdong, China
[4] Beckman Institute, University of Illinois at Urbana-Champaign, Urbana, IL, USA
[5] Department of Electrical and Computer Engineering, University of Illinois at Urbana-Champaign, Urbana, IL, USA

[§] These authors contributed equally to this work.

*Corresponding Authors:

    Dr. Shigao Chen
    Department of Radiology
    Mayo Clinic
    200 First Street SW
    Rochester, MN 55905
    Email: Chen.Shigao@mayo.edu

    Dr. Rongqin Zheng
    Department of Ultrasound
    Guangdong Key Laboratory of Liver Disease Research
    Third Affiliated Hospital of Sun Yat-Sen University
    600 Tianhe Road, Guangzhou 510630, China
    Email: zhengrq@mail.sysu.edu.cn

**Competing Interests:** The authors declare no competing interests.

**Funding:** The study was supported by the National Institute of Diabetes and Digestive and Kidney Diseases under R01DK120559, and by the National Cancer Institute under R00CA214523. The content is solely the responsibility of the authors and does not necessarily represent the official views of the National Institutes of Health.

**Abbreviations:**

ULM = ultrasound localization microscopy, MB = microbubble, CEUS = contrast-enhanced ultrasound, HFR = high frame-rate

# Abstract

**Background:**

Non-invasive detection of microvascular alterations in deep tissues *in vivo* provides critical information for clinical diagnosis and evaluation of a broad-spectrum of pathologies. Recently, the emergence of super-resolution ultrasound localization microscopy (ULM) offers new possibilities for clinical imaging of microvasculature at capillary level. Currently, the clinical utility of ULM on clinical ultrasound scanners is hindered by the technical limitations, such as long data acquisition time, high microbubbles (MBs) concentration, and compromised tracking performance associated with low imaging frame-rate. Here we present a robust in-human ULM on a high frame-rate (HFR) clinical ultrasound scanner to achieve super-resolution microvessel imaging using a short acquisition time (< 10s). Ultrasound MB data were acquired from different human tissues, including a liver, a kidney, a pancreatic tumor, and a breast mass using an HFR clinical scanner. By leveraging the HFR and advanced processing techniques including sub-pixel motion registration, MB signal separation, and Kalman filter-based tracking, MBs can be robustly localized and tracked for successful ULM under the circumstances of relatively high MB concentration associated with standard clinical MB administration and limited data acquisition time in humans. Subtle morphological and hemodynamic information in microvasculature were demonstrated based on data acquired with single breath-hold and free-hand scanning. Compared with contrast-enhanced power Doppler generated based on the same MB dataset, ULM showed a 5.7-fold resolution improvement in a vessel based on a linear transducer, and provided a wide-range blood flow speed measurement that is Doppler angle-independent. Microvasculatures with complex hemodynamics can be well-differentiated at super-resolution in both normal and pathological tissues. This study demonstrated the feasibility of ultrafast in-human ULM in various human tissues based on a clinical scanner that supports HFR imaging, and showed a great potential for the implementation of super-resolution ultrasound microvessel imaging in a myriad of clinical applications involving microvascular abnormalities and pathologies.

# Introduction

Microvasculature plays a critical role in blood supply for effective oxygen and nutrient delivery and metabolic demand of the tissue. Abnormal alterations of microvasculature are linked to a broad spectrum of diseases, such as arteriolosclerosis, diabetes, diabetes, chronic kidney diseases, hepatic fibrosis and cirrhosis, inflammatory bowel disease, and cancer (1-5). For example, diabetes is known to be related to various long-term microvascular complications, such as retinopathy and nephropathy, which lead to structural and functional vascular changes, eventually resulting in organ damage (2). The complications may have emerged in the level of small vessels long before they manifest as a symptom in the level of organs. Pathological angiogenesis, a hallmark of cancer, is another characteristic disease state that is associated with unbalanced development of microvasculature (e.g., chaotic, tortuous, and disordered small vessels that cause impaired blood flow) (1, 3). As such, early detection and characterization of the pathological changes of the microvasculature are of significant clinical value for disease prevention, intervention, and treatment. Despite the significance of microvasculature detection, there is no viable clinical tool that provides non-invasive and sufficient resolution for *in vivo* assessment of human tissue microvasculature (e.g., < 100 µm) at a clinically relevant depth (several centimeters to > 10 centimeters) (6, 7). *Ex vivo* histopathology remains the only clinical tool and gold standard for evaluating the state of health for tissue microvasculature. Current clinical modalities for vasculature imaging such as magnetic resonance imaging (MRI) and computed tomography (CT) can assess tissue perfusion or vascular structure with high sensitivity and deep penetration, but the spatial resolution is limited to the level of submillimeter or millimeter (> 100 µm) (8). Doppler ultrasound is another widely available modality for blood flow imaging. However, conventional ultrasound Doppler is only sensitive to fast flow (for example, > 1 cm/s) in relatively large vessels, and is challenging for imaging microvessels (6). Contrast-enhanced ultrasound (CEUS), which is routinely used in clinical practice for tissue perfusion imaging, allows a much higher sensitivity to flow in small vessels (from capillaries to tens of micrometers) by leveraging the microbubbles (MBs) that pass through the capillaries as an ultrasound contrast agent (9). However, CEUS is still not capable of resolving microvessels because it is confined by the same resolution limit as conventional ultrasound, which is approximately equal to the wavelength of ultrasound (e.g., from 100 um at 15 MHz to 1000 um at 1.5 MHz).

Recently, the invention of super-resolution ultrasound localization microscopy (ULM) achieved a substantial improvement of ultrasound spatial resolution in imaging microvasculature (6, 7, 10). Instead of using the diffraction-limited contrast microbubble (MB) signal for imaging as in contrast-enhanced ultrasound (CEUS), ULM uses the location of the spatially isolated MBs for imaging. ULM has demonstrated an approximately ten-fold improvement in spatial resolution while preserving the imaging penetration of conventional ultrasound (6, 7, 10). In addition, movement of individual MBs can be tracked to provide an accurate measurement of a wide range of blood flow speed that is Doppler angle-independent. Owing to the significant clinical potential, ULM has rapidly gained traction in the research field with focus on the technical development, optimization, and demonstration (11-30), which includes various phantom validation studies (17, 19, 31-35) and a wide range of preclinical *in vivo* studies in different tissues (e.g., brain, tumor, kidney, ear, skeletal muscle, lymph node) of various animal models including mouse (11, 12, 14), rat (10, 13, 15, 30, 36-38), rabbit (18, 22, 26), and chicken embryo (16, 39).

As an emerging technology, most of the ULM implementations to date are still limited to research ultrasound scanners with high frame-rate (HFR) imaging capability. There are pilot in-human studies in imaging breast tumors (11, 40), lower limb (41), and prostate cancer (42) using conventional clinical scanners that typically operated at low imaging frame-rates (around 10 to 15 Hz). However, it is challenging to accurately localize and track fast moving MBs with a low frame-rates, which compromises the performance of the ULM. Although MB dilution is commonly used to mitigate the issue, it is not a standard clinical procedure for MB administration. Another challenge with conventional ULM is the long data acquisition time, which requires the same imaging plane to be continuously imaged for dozens of seconds and even several minutes (11, 40-42). In practice this is very challenging for free-hand scanning, especially for organs like kidney and liver where tissue motion from patient breathing is significant and difficult to manage. While patient breath-holds mitigate the issue of tissue motion, it is infeasible to suspend voluntary respiration over the entirety of ULM data acquisition.

To address these limitations of ULM, we recently developed an MB signal separation method and a Kalman filter-based tracking algorithm to facilitate accurate localization and tracking of MBs without the need of MB dilution. Our methods also significantly shorten ULM data acquisition time to only several seconds (39, 43). In this paper, we take advantage of a clinical scanner with the capability of HFR imaging (e.g., > 500 Hz) and implement these algorithms to realize robust in-human ULM with a short acquisition time (< 10 s). We applied the MB signal separation method on the clinical HFR ultrasound system to enable accurate localization and tracking of MBs using standard clinical MB administrations without MB dilution(39). We further improved microvessel delineation by implementing the Kalman filter-based tracking algorithm on the clinical HFR system (43). We demonstrated successful in-human ULM using data acquired from a single breath-hold and free-hand scanning. The utility of the technique was tested in various *in vivo* human tissues including liver, kidney, and pancreatic and breast tumors. The results demonstrated a 57.5 µm spatial resolution and an imaging penetration depth of up to 8 cm using a linear array transducer (3.6 MHz). To the best of our knowledge, this is the first in-human ULM imaging using a clinical scanner that supports HFR imaging. With the trend of more and more clinical ultrasound scanners supporting HFR imaging, our study demonstrated the feasibility of implementing ULM on these scanners and paved the way for future clinical translations of the ULM technology.

# Methods

## *Study Protocol*

This study was approved by the Institutional Review Board (IRB) of the Third Affiliated Hospital of Sun Yat-Sen University and the IRB of Mayo Clinic. Written informed consent was obtained from each patient. For each patient, a bolus of MB solution (SonoVue, Bracco, Milan, Italy) was injected through intravenous injection, followed by a 5 ml saline flushing. The dosage of MB was organ-specific, similar to that in clinical CEUS (see details in Table I). Real-time CEUS was used to guide the image plane and monitor the MB signal right after the MB injection with free-hand scanning. Then the sonography switched the scanner to an ultrafast plane-wave imaging mode and started the data acquisition. The beamformed in-phase/quadrature (IQ) data were saved for ULM post-processing in MATLAB (MathWorks Inc., Natick, MA) offline. During the data acquisition, the patients were asked to hold the breath for a data acquisition time < 10 seconds. The data were acquired from different tissues, including a liver, a kidney, a pancreatic and a breast tumor from different patients.

## *Ultrasound Imaging Settings*

An HFR clinical ultrasound scanner (Resona 7, Shenzhen Mindray Bio-Medical Electronics Co. Ltd., Shenzhen, China) and an L9-3U linear array transducer (6.0 MHz center frequency, bandwidth 9.8MHz – 1.8MHz) were used for both real-time CEUS and ULM data acquisition. Ultrafast imaging was performed using a 5-angle plane-wave compounding (-10° to 10°, 3.6 MHz frequency, 1.5 cycles of pulse length) with a compounded frame-rate of several hundred Hertz (depending on the image depth, Table 1). A stack of several thousands of ultrasound frames (Table 1) were acquired for each tissue for the given acquisition time and frame-rate.

## *Ultrasound signal processing*

A spatiotemporal singular-value-decomposition (SVD) based clutter filter was applied to the ultrasound data (Fig.1a) to extract moving MB signals (Fig. 1b, corresponding to the median-to-high-rank components) (44, 45). The MB signals can be used to generate contrast-enhanced power Doppler images by accumulating the power of the MB signals along the temporal dimension. It should be noted that this power Doppler image is based on the contrast MB signals and thus different from the non-contrast conventional power Doppler image. The same MB data will be used for localization and tracking in ULM. In this study, a phase correlation-based sub-pixel motion estimation method was applied to estimate tissue motion in 2-D based on the IQ data (with MB signals removed by the SVD filter) (43, 46), which will be utilized in the following steps to register the MB signals for robust tissue motion correction. By separating the MB signals and tissue signals, the influence of the moving MB signals on the tissue motion estimation can be suppressed (47).

## *MB localization and tracking for ULM*

An MB signal separation method was applied to break down the original MB data into several subsets of low MB concentration data based on MB flowing speeds and directions (39). With MB separation, the spatially overlapping MBs that are otherwise discarded in the conventional ULM processing may now be utilized for robust localization and tracking separately in each subset, to improve the efficiency and accuracy of MB event detection for the limited acquisition time (39). For each subset, tissue motion was removed based on the above motion estimation. The motion-corrected MB signals were equalized by a noise profile and then spatially interpolated to an axial-lateral pixel resolution of 57.5 µm using a 2-D spline interpolation. An intensity thresholding was applied to remove the noisy background. A 2-D normalized cross-correlation between interpolated MB data and a predefined point-spread function (PSF) was calculated, followed by a thresholding to remove pixels with correlation coefficient < 0.6. MB centroids were localized as the regional peaks of the cross-correlation map (Fig. 1b). A bipartite graph-based pairing algorithm was applied to track the movement of individual MBs over frames. Only those MBs consistently tracked over ten compounded ultrasound frames were preserved as confident MB events for generating microvessel images. Each MB trace was smoothed and the gaps in between MB positions on each MB trace were inpainted to further improve the microvessel delineation using a Kalman-filter-based algorithm (43). An MB movement direction and acceleration constraint were used to further remove noisy MB traces (43), before accumulation of the MB traces from all the subsets to generate the final microvessel images (39).

# Results

## *Robust ULM with Ultrafast imaging*

The accumulation of MB positions from all the subsets generates an MB density image that depicts the vessel morphology (Fig. 1c, without motion registration). The Kalman filter-based algorithm was applied to smooth and inpaint the MB traces (43), further improving ULM image quality (Fig. 1e). Fig. 1d shows an example of a lateral motion curve (blue line). Even with breath-holding, tissue motion was still present (200 µm – 400 µm), which resulted in deteriorated microvessel imaging (Fig. 1e). Here a robust phase correlation-based motion correction method based on ultrafast ultrasound imaging was used to remove the tissue motion. The residual tissue motion was reduced to < 15 µm (red line, Fig. 1d), which falls well below the pixel size of the ULM image (57.5 µm × 57.5 µm). The motion-corrected microvessel image is depicted in Fig. 1f, showing an improved resolution.

## *ULM of Healthy Human Liver Microvasculature*

The ULM images of a human liver with a field-of-view (FOV) of 58 mm × 44 mm (depth × width) derived from 6.0 s of data (415.3 Hz frame-rate) are shown in Fig. 2. Fig. 2b shows the zoomed-in region depicting the detailed microvessel structures with a wide range of vessel diameters (millimeters to tens of micrometers). To further highlight the ULM resolution, a smaller region (indicated by the rectangle in Fig. 2b) was magnified (Fig. 2c), and compared with contrast-enhanced power Doppler image generated using the same HFR dataset as those used for ULM (Fig. 2d, see Sub. Fig. 1a for a larger FOV image). There are some discontinuities of the vessel profile inside the relative large vessel indicated by the arrow in Fig. 2c, which may be associated with the limited number of MB tracks detected in a short acquisition time. Vessel profile along the dashed line in Fig. 2c indicates a vessel width of 153 µm (full-width at half maximum, FWHM), showing about a 5.7-fold improvement of spatial resolution compared with the contrast-enhanced power Doppler image (869 µm, Fig. 2e) for the current imaging settings. Two branching vessels at the bifurcation with a width of 158 μm and 105 μm (575 μm distance), as indicated in Fig. 2e (right side), can be clearly resolved, but were inseparable in the power Doppler. Fig. 2f shows the bi-directional microvessel intensity image, similar to Fig. 2b but with red color indicates the upward flow and blue color represents the downward flow. Reasonable gradients of flow speeds from larger vessels to smaller vessels can be observed, as evidenced in the velocity magnitude image in Fig. 2g (see the bi-directional version of microvessel velocity image in Sub. Fig.1c). A wide range of flow speeds (from about 1 mm/s to about 60 mm/s) can be measured as indicated by the distribution of the velocity magnitude (Fig. 2h).

### *ULM of Healthy Human Kidney*

The cortex of the human kidney represents a complex flow pattern with high perfusion, dense microvascular structure. A B-mode image of the kidney was shown in Sup. Fig.2a, indicating the orientation of the image plane. MB separation was able to separate MBs with opposite flow directions in cortical arteries and veins, thus yielding a higher number of high-fidelity MB traces for successful ULM with a short acquisition time (9.7 s of data with 250.4 Hz frame-rate), as shown in Fig. 3. The super-resolution microvessel intensity, bi-directional intensity and velocity magnitude images are shown in Fig. 3a-3c, respectively. Well-defined microvessel structure can be resolved, showing clear vessel boundaries in the cortex region, which are difficult to discern in the power Doppler image (Sup. Fig. 2b). Fig. 3d and 3e show the magnified region of the bi-directional intensity and velocity magnitude images at the upper cortex of the kidney, showing the details of the dense microvasculature and complex hemodynamics. In the bi-directional intensity image (Fig. 3d), a pixel is either assigned to red or blue colors even though there may be several vessels with opposite flow directions passing thought the same pixel, which can lead to the 'aliasing-like' artifact indicated by the arrows in Fig. 3d. This can also easily appear in the horizontal vessel where flow direction is close to zero (or 180°) and pixels are hard split into red and blue colors based on the sign of the flow direction. Flow directions and the neighboring arteries and veins with opposite flows can be well-differentiated (Fig. 3d, and Sup. Fig. 2c-2d).

## ULM of In-human Pancreatic Tumor and Breast Tumor

We tested the ULM in a human pancreatic tumor with a depth down to 60 mm, and a shallow breast tumor down to 15 mm (Figs. 4-5). The pancreatic tumor data was acquired from a 70-year-old patient with an acquisition time of 5.2 s (frame-rate 415.3 Hz). This tumor had a large size with a dimension of about 164 mm × 65 mm × 66 mm with an anterior lesion boundary at a depth of around 10 mm, as assessed from the clinical ultrasound and CT scan. The acquired ultrasound data can only cover a portion of the lesion, as indicated in Fig. 4. The B-mode ultrasound indicates a mixed cystic solid mass with a dimension of 50 mm × 50 mm in the image FOV (with the boundary of the tumor was indicated by white arrows in Fig. 4a). ULM showed both the morphological and hemodynamic characteristics with a high spatial resolution (Fig. 4b- 4d). Figs. 4e-4f shows the magnified region of the power Doppler, ULM intensity, and velocity magnitude images at a deeper region (40 to 55 mm), respectively, revealing substantially improved resolution by the ULM. Feeding and draining vessels with opposite flow directions can be well-differentiated in the bi-directional intensity image (Fig. 4c, indicated by white and yellow arrows, respectively) or in the flow direction image and bi-directional velocity image (Sup. Fig. 3b and 3c).

The B-mode ultrasound of the breast tumor is shown in Fig. 5a, with white arrows indicating the boundary of the lesion. The data was acquired from a ductal carcinoma in situ (DCIS) diagnosed with histopathology on a 49-year-old patient who had been treated with neoadjuvant chemotherapies (9.0 seconds of data, frame-rate at 622 Hz). Similarly, substantial improvement in resolution was observed for ULM (Figs. 5b-5d) over the power Doppler (Sup. Fig. 4a). Local regions of power Doppler, ULM density and velocity magnitude images are shown in Figs. 5e-5g, respectively, for detailed visualization.

## Discussion

Conventional ULM requires a long data acquisition time (dozens of seconds to minutes) to accumulate a sufficient amount of isolated MB signals for full delineation of the microvasculature. With the combination of HFR ultrasound imaging and advanced post-processing techniques (MB separation, Kalman-filter tracking, and sub-pixel motion registration) presented in this paper, robust ULM can be accomplished *in vivo* in humans using standard clinical procedure of MB administration and short data acquisition with a single breath-hold (< 10s), and free-hand scanning. In this study, we demonstrated the feasibility of implementing fast and robust ULM on a clinical ultrasound scanner that is capable of HFR imaging. We successfully performed ULM imaging in various human tissues *in vivo*. The results showed well-defined morphological and functional features in the microvasculature in both healthy (liver, kidney) and pathological tissues (pancreatic and breast tumors). Given that more and more clinical ultrasound scanners support the HFR imaging, there is a great potential for the methods presented in this paper to be widely distributed to enhance the clinical accessibility of ULM.

Contrast-enhanced power Doppler image derived from the same dataset used for ULM processing was utilized as a benchmark for resolution comparison in this study. In liver vessels with clear vascular branches at a depth of around 35 mm, we showed an about 5.7-fold resolution improvement of ULM compared with the contrast-enhanced power Doppler image. It should be noted that this improvement evaluation was based on one single vessel (Fig. 2c-2e), which has a vessel width of 153 µm (FWHM) measured in ULM and 869 µm measured in power Doppler image, and does not represent the highest resolution that can be achieved by either ULM or power Doppler image. In fact, contrast-enhanced power Doppler image, similar to the conventional ultrasound, can potentially achieve an axial resolution of approximately half-wavelength to one-wavelength, which is about 214 µm to 428 µm for the 3.6 MHz center frequency. This spatial resolution by conventional ultrasound image can be increased by using a higher ultrasound frequency, but still limited ultrasound diffraction. Furthermore, a lateral profile was used to measure the vessel width here (Fig. 2e, 869 µm), which was related to the lateral resolution of the ultrasound image and was expected to be much worse than the ultrasound resolution in the axial direction. Nevertheless, by using the same data set under the same imaging settings, the ULM is shown to be capable of imaging vasculature with several-fold resolution improvement *in vivo*.

In this study, the super-resolved microvessel images were reconstructed at a 57.5 µm axial/lateral pixel resolution. This spatial resolution can be further improved by using a smaller pixel size down to about 1/20 to 1/10 of the acoustic wavelength (that is, 21 µm to 43 µm for a 3.6 MHz imaging frequency). However, a much longer acquisition time will be required to fully populate the microvasculature at such fine resolutions (6, 48-50). The required acquisition time for imaging capillary flows has been thoroughly evaluated by Christensen-Jeffries *et al* (49), Hingot *et al (48)*, and Lowerison *et al (50)*. that an extremely long time (e.g., tens of minutes) would be necessary for the capillary networks to be fully filled by MBs. Therefore, it is a physiological limitation of the technique that it would not be possible to image the capillary networks at the level of micrometers in less than 10 s used in this study. While increasing the MB concentration can speed up capillary perfusion of MBs, it also increases the challenge to obtain isolated MB signals for ULM imaging with elevated MB concentration even with the MB separation technique. Therefore, this is a tradeoff between the resolution of microvessel imaging and clinically practical condition of data acquisition time. As such, the inability to resolve full capillary bed remains a limitation for in-human ULM. Furthermore, the accuracy of motion registration will also limit the capability of imaging capillaries in human *in vivo* even a long acquisition is applied, especially for the liver or kidney, where tissue motion is significant and inevitable. However, for clinical applications such as liver and kidney the velocities in the images are mainly in the range of tens of millimeters per second, which are flows 10-20 times faster than that in the capillary networks. This level of vessels is supposed to be filled by MB much faster (may be tens of times faster) than capillaries with the slower flow. Based on our results in this study, it 's possible to image vessels smaller than 100 μm using several seconds of data acquisition and using clinical relevant MB concentrations.  Except for the practical condition for clinical scanning, the clinical needs for the imaging resolution of in-human ULM may also need to be considered. Whether the capability of resolving vessels at the level of arterioles and venules is adequate for clinical assessment of microvascular pathologies remains to be investigated.

Except for morphology imaging of the microvasculature, another advantage of ULM is the capability of providing 2-D flow velocity estimation that is Doppler angle-independent by tacking the individual MB movement. However, it typically requires much longer acquisition time (e.g., 4 times longer) to reconstruct well-developed velocity profiles inside the vessels than the time required for reconstructing a microvessel intensity image (48). For each MB track, instantaneous MB moving velocity can be accurately measured, and the final velocity image is generated by averaging all velocity estimates of individual MBs passing through the same pixel. The instantaneous moving speed of different detected MBs pass through the same vessel or pixel can be very different due to the hemodynamic change associated with the cardiac phase, and the poor elevation resolution. It may only need at least one MB to reconstruct a pixel of a vessel for the morphology image. However, to accurately estimate the velocity profiles, multiple MBs passing through the same pixel will be necessary for a stable averaged velocity estimate. Therefore, even morphology image can be very well-reconstructed, the number of MBs detected may not be sufficient to provide smooth velocity estimation in some of the microvessels, which is an inherent limitation of the ULM based on short acquisition times. As a consequence, there are some discontinuities and variations of velocity profiles observed in the velocity images in this study. There are possibilities that different velocities appear in one single vessel or a relatively larger vessel shows a slower flow than the tiny branches. This issue of undersampling of velocity profiles would be potentially improved with more MB detections using a longer acquisition time. Nevertheless, the flow direction is less related to the cardiac phase and can be accurately estimated even there are only a few MBs passing through one pixel, which will provide added information combined with the morphology image, such as the bi-directional intensity image depicting both flow direction information and morphological information of the microvasculature.

A relatively large FOV and clinical relevant penetration depths were demonstrated for ULM in this study. However, since a linear array transducer was used, the optimal imaging performance was found mainly in depth less than about 5 cm. For some applications such as imaging deeper abdominal tissues (> 5 cm), a curved array or phased array transducer operating at lower frequencies can be deployed. As a pilot study, only a linear array transducer was used here to demonstrate in-human ULM, and the same ultrasound frequency (3.6 MHz) was used here for all of the organs from the shallowing breast tissue to the deeper kidney tissue for the purpose of concept proofing. However, the methods presented in this study can be conveniently translated to other transducers at different frequencies optimized for different applications. One limitation of the study is that there was no ground truth for validating the microvasculature imaged by ULM. This is a common challenge for *in vivo* studies. Nevertheless, we were able to obtain physiologically reasonable microvascular images with well-defined vessel network. The ULM requires a certain period of acquisition time (< 10 s in this study) to collect data of a sufficient number of MBs filling in the vasculature, which prohibits it from being a real-time imaging modality. Moreover, post-processing of a large amount of ULM data associated with long acquisition time, including SVD clutter filtering, MB localization, pairing, and tracking, is still time-consuming, which, however, has the possibility to be speeded up via parallel processing using high power GPU. In addition, two-dimensional (2-D) ULM suffers from inaccurate blood flow velocity estimation even with very long acquisition time due to the tracking of 3-D MB movements using a 2-D imaging plane. The 2-D ultrasound image casts a slice thickness in the elevational direction (perpendicular to the 2-D image plane), which renders traces of MBs that are moving in-and-out-of-plane. 3-D ULM is necessary to accurately measure blood flow speed in all three dimensions. Additionally, 3-D imaging can also enhance the performance of motion correction by providing 3-D motion estimation.

In conclusion, we demonstrated successful in-human ultrasound localization microscopy (ULM) imaging using an HFR clinical ultrasound scanner, and showed high-definition morphological and functional imaging of microvasculature at super-resolution. By leveraging the high imaging speed and advanced post-processing algorithms, in-human ULM can be accomplished using data acquired from a single breath-hold (< 10 s) and free-hand scanning. Successful ULM imaging was demonstrated in various *in vivo* human tissues including liver, kidney, and pancreatic and breast tumors, showing the in-human feasibility of the technique. The results of the study show great potential for future clinical translations of ULM based on clinical HFR scanners that are rapidly becoming available around the world.

# Tables

TABLE I
ULTRASOUND IMAGING SETTINGS FOR SPECIFIC TISSUES IN THIS STUDY

|  | Frame-rate (Hz) | Data acquisition time (s) | Number of ultrasound frames | Bolus injection (ml) |
|---|---|---|---|---|
| Liver | 415 | 6.0 | 2487 | 2.4 |
| Kidney | 250 | 9.7 | 2424 | 2.0 |
| Pancreatic tumor | 415 | 5.2 | 2144 | 2.0 |
| Breast tumor | 622 | 9.0 | 5575 | 3.0 |

# Figure Legends

**Figure 1.** (a) Stack of B-mode images indicating original spatiotemporal ultrasound data from a healthy human liver (t represents slow time). (b) B-mode images of the MB signal extract from (a) with tissue clutter filtering. Centroids of MBs were localized and indicated by the red asterisks obtained from the cross-correlation with a predefined point-spread-function (PSF) of the system. (c) Accumulation of the MB centroid positions over time to generate an MB density map. (d) The estimated lateral tissue motion curve with respect to the first ultrasound frame before and after motion registration. (e) the microvessel density map obtained from accumulating MB positions with MB pairing, tracking, Kalman filtering, but without motion correction. (f) The microvessel density map with motion correction. Here, the example images in (c)(e)(f) were generated with 829 frames of MB data, corresponding to a data acquisition time of 2.0 s.

**Figure 2**. (a) Full FOV super-resolution ULM microvessel image of the healthy human liver overlaid on the B-mode image. (b) The zoomed-in local region as indicated by the white rectangle in (a).. (c) The magnified region indicated by the white rectangle in (b) highlighting a small branch of vessels. (d) The contrast-enhanced power Doppler image of the same branch of vessels in (c), which is obtained by accumulating the power of the MB signal over time. (e) Plots of a vessel profile along the white dashed line indicated in (c) using ULM (red curve)and power Doppler (blue curve). (f) Bi-directional super-resolution microvessel intensity image, similar to Fig. 2b but with red color indicates the upward flow and blue color represents the downward flow. (g) Super-resolution ULM microvessel velocity image of the healthy human liver, with colorbar indicating the magnitude of the velocity. The direction of the blood flow can be found in Sup. Fig. 1. (h) The histogram of the blood flow speed distribution.

**Figure 3**. (a) Super-resolution microvessel intensity image of a human kidney overlap on the ultrasound B-mode image. (b) corresponding bi-directional microvessel intensity image, similar to Fig. 3a, but with red color indicates the upward flow and blue color represents the downward flow. Neighboring cortical arteries and veins with opposite flow directions can be well-differentiated at high spatial resolution. (c) Corresponding super-resolution microvessel velocity image of the human kidney, with colormap indicating the magnitude of the velocity. The direction of the blood flow can be found in Sup. Fig. 2(b). (d) Zoom-in region of the bi-directional super-resolution microvessel intensity image and (c) Zoom-in region of super-resolution microvessel velocity image indicated by the white rectangle in (c) showing dense microvasculature and complex hemodynamics in the renal cortex.

**Figure 4**. (a) B-mode image of a pancreatic tumor from a 70-year-old patient, with white arrows roughly indicating the boundary of the lesion. (b) Super-resolution microvessel intensity image of the human pancreatic tumor. (c) Corresponding bi-directional microvessel intensity image, similar to Fig. 4b but with red color indicates the upward flow and blue color represents the downward flow. (d) Corresponding super-resolution microvessel velocity image, with colormap indicating the magnitude of the velocity. The direction of the blood flow can be found in Sup. Fig. 3b. (e) Zoom-in region (indicated by the white rectangle in Fig. 4b) of the contrast-enhanced power Doppler image. (f) Corresponding zoom-in region of the super-resolution microvessel intensity image. (g) Corresponding zoom-in region of the super-resolution microvessel velocity image.

**Figure 5**. (a) B-mode image of a breast tumor diagnosed as ductal carcinoma in situ from a 49-year-old patient after neoadjuvant chemotherapies, with white arrows roughly indicating the boundary of the lesion. (b) Corresponding super-resolution microvessel intensity image of the human breast tumor. (c) Corresponding bi-directional microvessel intensity image, similar to Fig. 5b but with red color indicates the upward flow and blue color represents the downward flow. (d) Corresponding super-resolution microvessel velocity image, with colormap indicating the magnitude of the velocity. (e) Zoom-in region (indicated by the white rectangle in Fig. 4b) of the contrast-enhanced power Doppler image. (f) Corresponding zoom-in region of the super-resolution microvessel intensity image. (g) Corresponding zoom-in region of the super-resolution microvessel velocity image.

**Supplemental Figure 1**. (a) Contrast-enhanced power Doppler image of the healthy human liver derived from the same MB data used for ULM. (b) Super-resolution microvessel image with the color map indicating the flow direction, where 0° indicates the left-to-right flow, 90° indicates the upward flow, and -90° representing the downward flow motion. (c) Bi-directional super-resolution microvessel velocity image, with red color (positive) indicates the upward flow and blue color (negative) represents the downward flow.

**Supplemental Figure 2.** (a) B-mode image of the human kidney. (b) Contrast-enhanced power Doppler image of the human kidney derived from the same MB data used for ULM. (c) Super-resolution flow direction image of the human kidney with the color map indicating the flow direction, where 0° indicates the left-to-right flow, 90° indicates the upward flow, and -90° representing the downward flow motion. (d) Bi-directional super-resolution microvessel velocity image, with red color (positive) indicates the upward flow and blue color (negative) represents the downward flow.

**Supplemental Figure 3**. (a) Contrast-enhanced power Doppler image of the human pancreatic tumor derived from the same MB data used for ULM. (b) Super-resolution microvessel image of the human pancreatic tumor with the color map indicating the flow direction, where 0° indicates the rightward flow, 90° indicates the upward flow, and -90° representing the downward flow motion. (c) Bi-directional super-resolution microvessel velocity image, with red color (positive) indicates the upward flow and blue color (negative) represents the downward flow.

**Supplemental Figure 4**. (a) Contrast-enhanced power Doppler image of the human breast tumor derived from the same MB data used for ULM. (b) Super-resolution microvessel image of the human breast tumor with the color map indicating the flow direction, where 0° indicates the rightward flow, 90° indicates the upward flow, and -90° representing the downward flow motion. (c) Bi-directional super-resolution microvessel velocity image, with red color (positive) indicates the upward flow and blue color (negative) represents the downward flow.

**Figure 1.**

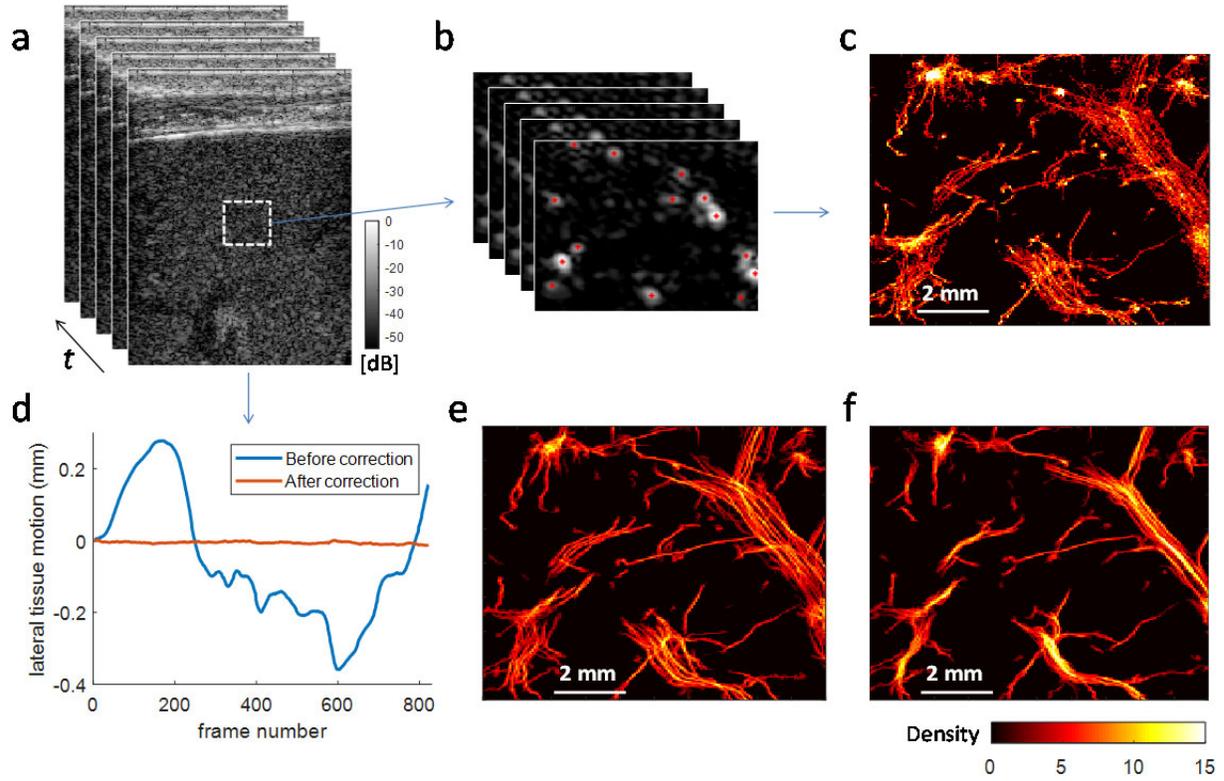

**Figure 2.**

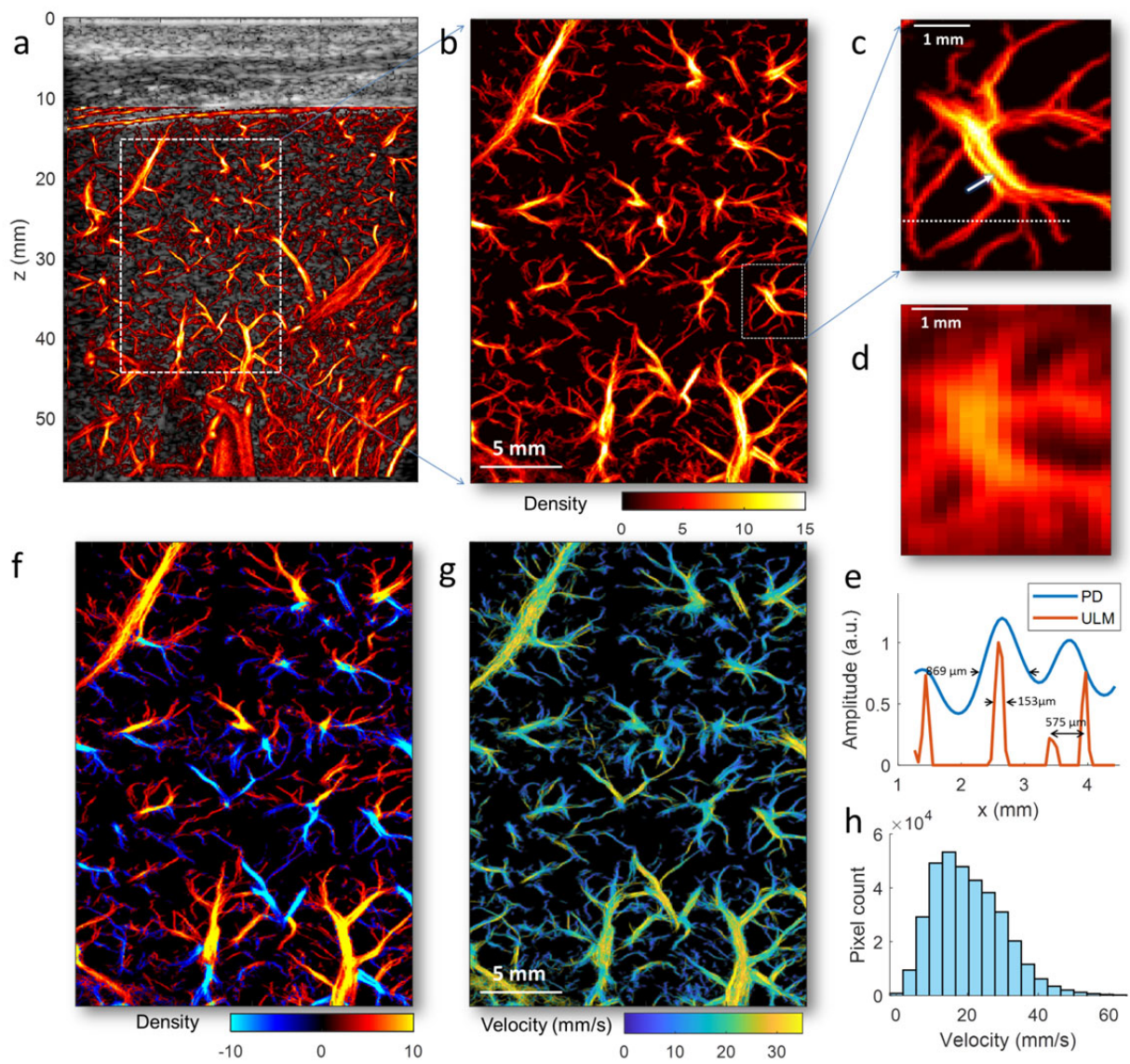

**Figure 3.**

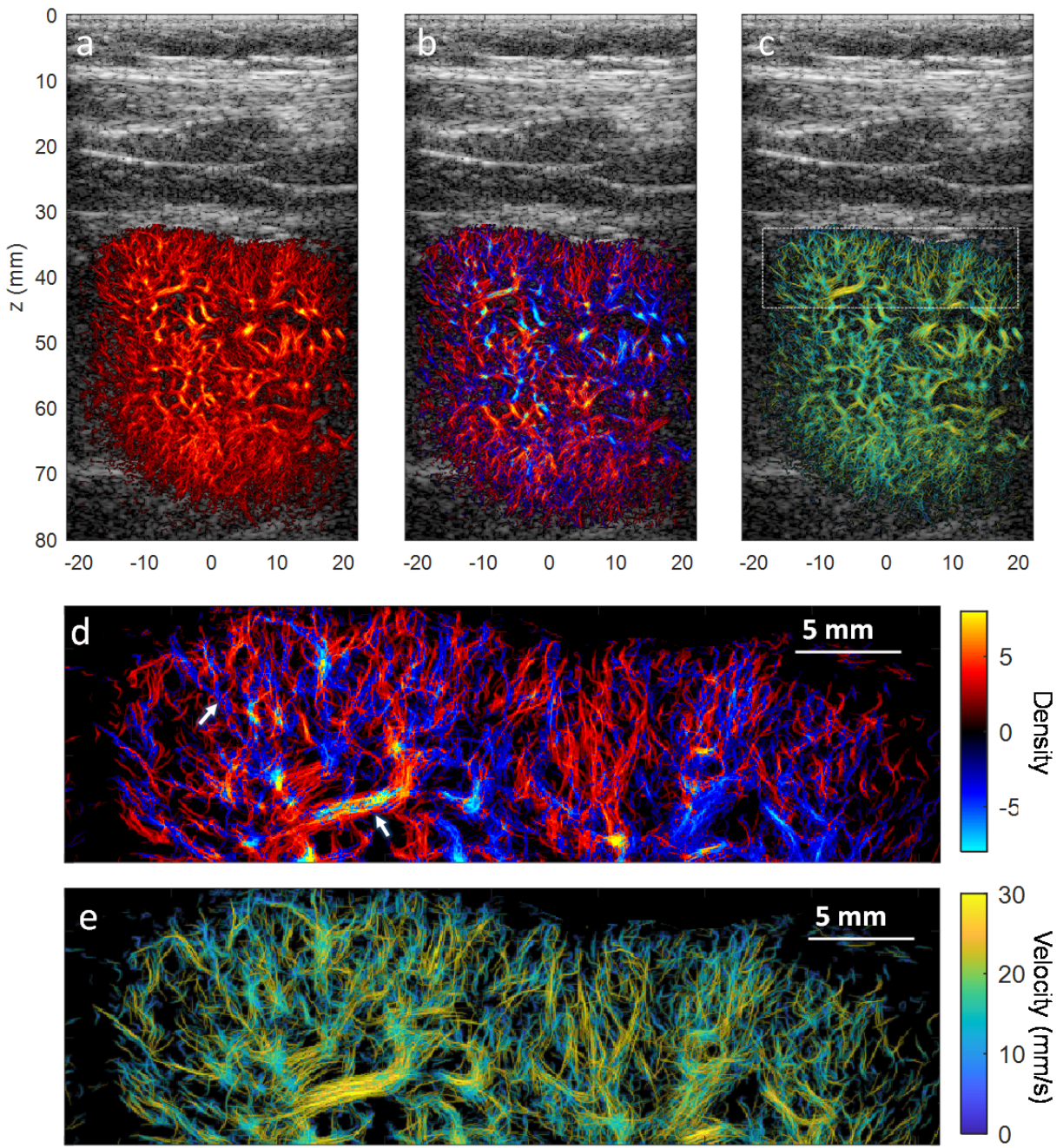

**Figure 4.**

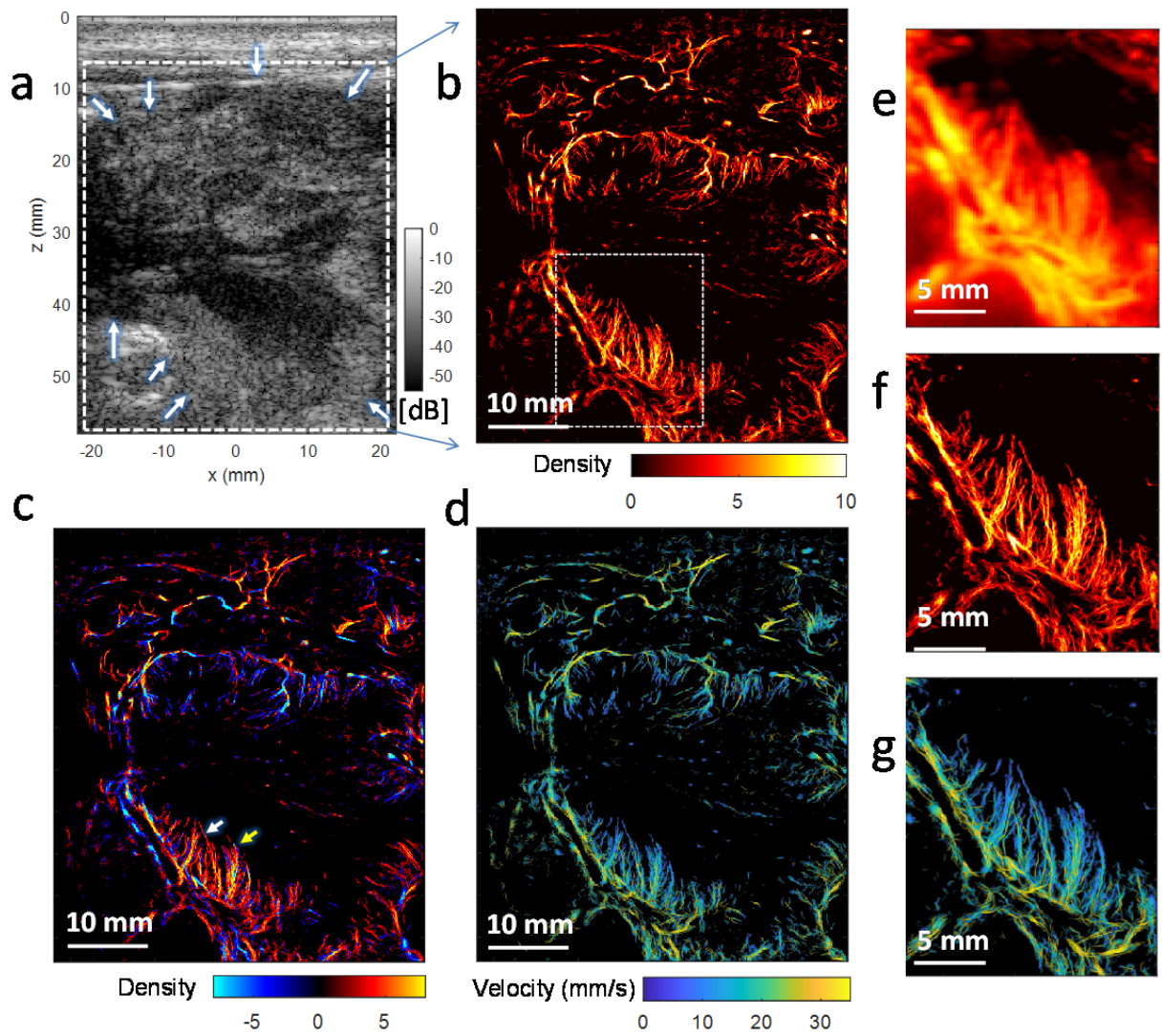

**Figure 5.**

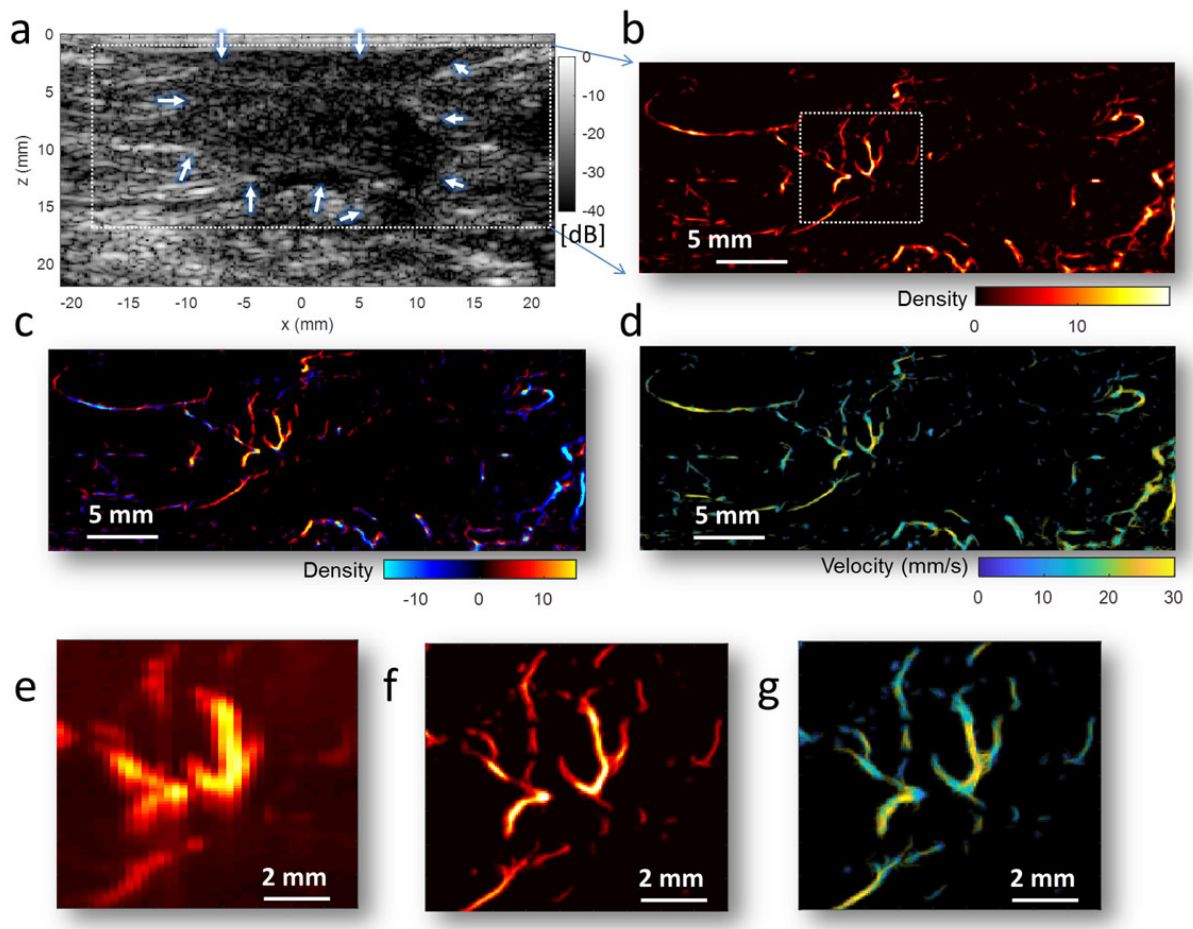

**Supplemental Figure 1.**

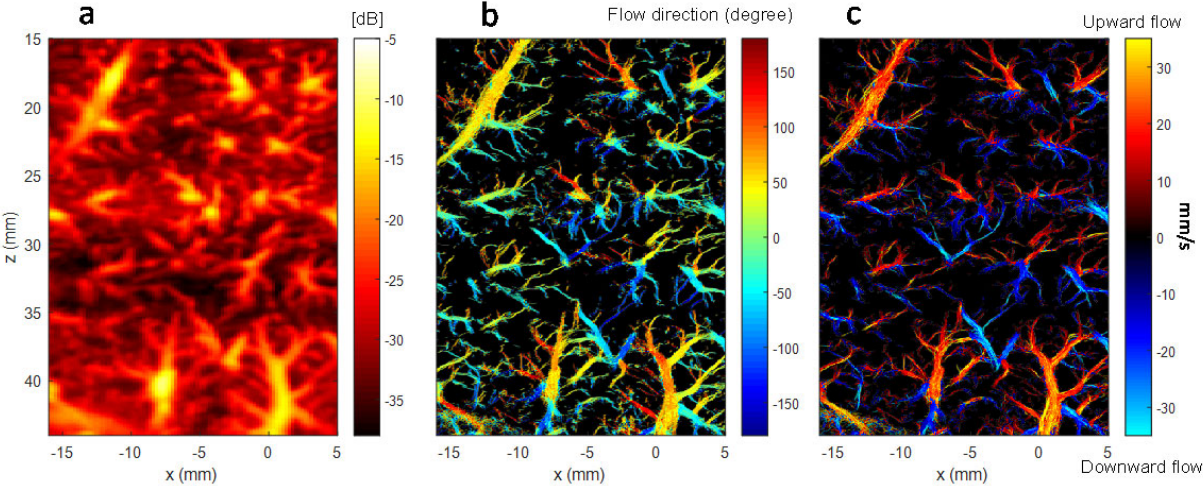

**Supplemental Figure 2.**

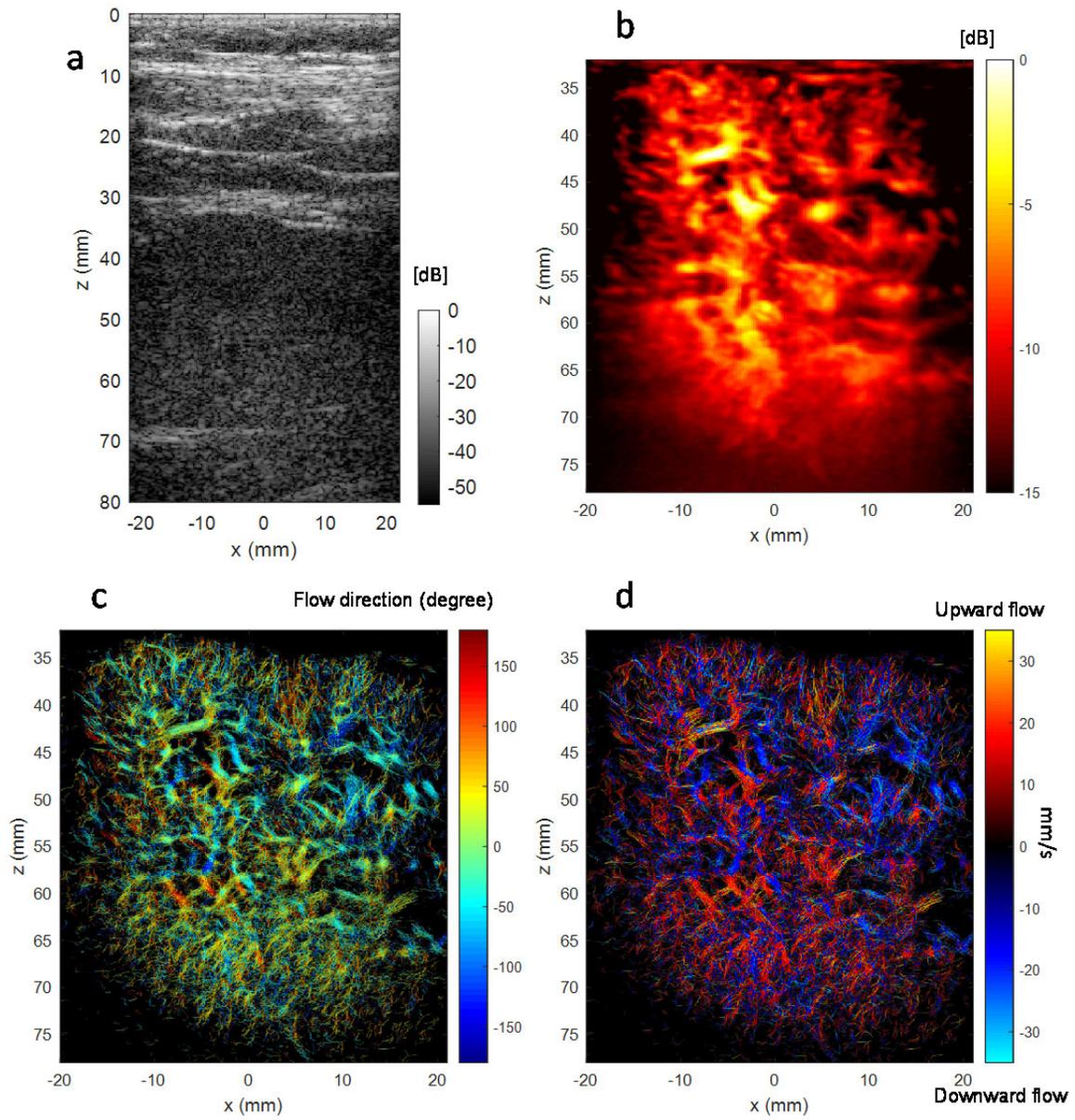

**Supplemental Figure 3.**

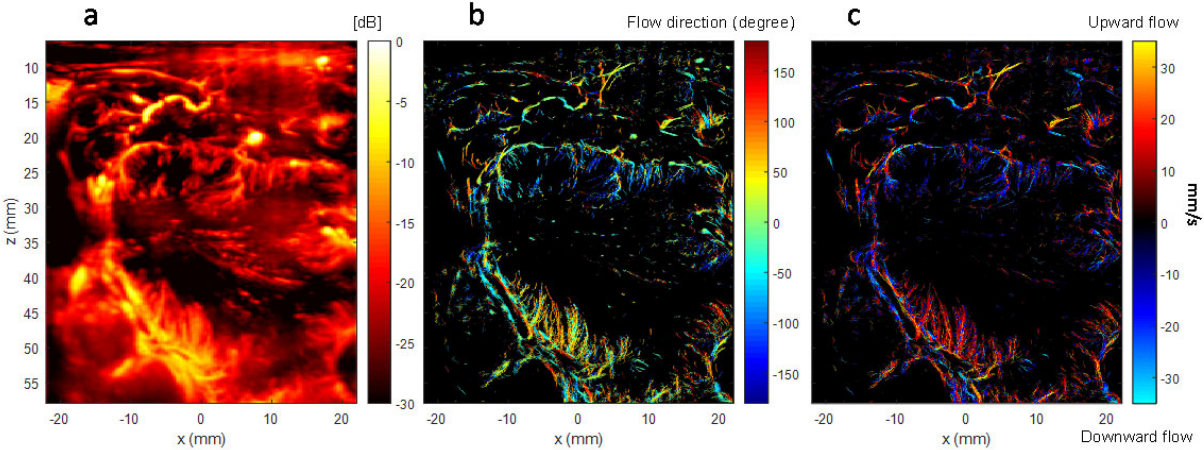

**Supplemental Figure 4.**

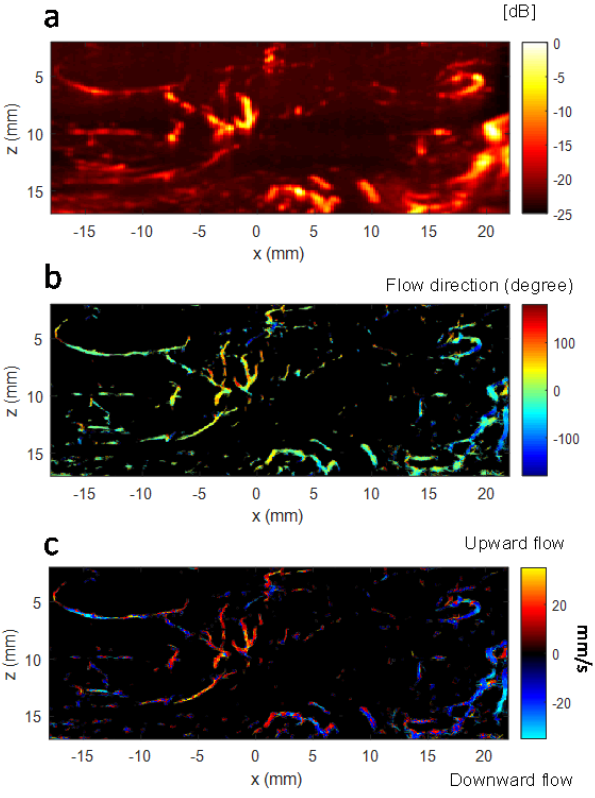